\documentclass[usenatbib]{mn2e}
\usepackage{graphicx, amssymb, aas_macros, natbib,array}
\newcommand{\vmax}{V_{\rm max}}

\newcommand{\mpeak}{M_{\rm{peak}}}

\newcommand{\mvir}{M_{\rm{vir}}}
\newcommand{\mhalo}{M_{\rm{halo}}}

\newcommand{\rvir}{R_{\rm{vir}}}

\newcommand{\mstar}{M_{\rm \star}}
\newcommand{\msun}{\rm \, M_{\odot}}

\newcommand{\lsun}{L_{\odot}}
\newcommand{\hmpc}{h^{-1} \, \rm Mpc}

\newcommand{\mpc}{\rm \, Mpc}
\newcommand{\kpc}{\rm \, kpc}
\newcommand{\pc}{\rm \, pc}
\newcommand{\gyr}{\rm \, Gyr}
\newcommand{\myr}{\rm \, Myr}
\newcommand{\kms}{{\rm km \, s}^{-1}}

\newcommand{\cmcb}{{\rm cm}^{-3}}

\newcommand{\msii}{MS-II}

\newcommand{\lcdm}{$\Lambda$CDM}

\newcommand{\dearly}{\rm Dwarf \, 2_{Early}}
\newcommand{\dmid}{\rm Dwarf \, 2_{Middle}}
\newcommand{\dlate}{\rm Dwarf \, 2_{Late}}
\newcommand{\done}{\rm Dwarf \, 1}
\newcommand{\dtwo}{\rm Dwarf \, 2}
\newcommand{\ufdone}{\rm UFD \, 1}
\newcommand{\ufdtwo}{\rm UFD \, 2}

\begin{document}

\title[Isolated ultrafaints and tiny satellites]
{
Sweating the small stuff: simulating dwarf galaxies, ultra-faint dwarf galaxies, and their own tiny satellites 
}\author[C. Wheeler et al.]{Coral Wheeler$^1$\thanks{$\!$crwheele@uci.edu},
Jose Onorbe$^{2}$, James S. Bullock$^1$, Michael Boylan-Kolchin$^{3}$ 
  \newauthor Oliver D. Elbert$^{1}$, Shea Garrison-Kimmel$^{1}$, Philip F. Hopkins$^{4}$, Dusan Keres$^{5}$\\
  \noindent$\!\!$ $^{1}$Center for Cosmology, Department of Physics and Astronomy,
  University of California, Irvine, CA 92697, USA \\
  \noindent$\!\!$ $^{2}$Max-Planck-Institut fuer Astronomie, Koenigstuhl 17, 69117 Heidelberg, Germany \\
    \noindent$\!\!$ $^{3}$Department of Astronomy and Joint Space-Science Institute,
    University of Maryland, College Park, MD 20742-2421, USA\\
    \noindent$\!\!$ $^{4}$TAPIR, Mailcode 350-17, California Institute of Technology, Pasadena, CA 91125, USA\\
    \noindent$\!\!$ $^{5}$Department of Physics, Center for Astrophysics and Space Sciences, University of California at San Diego, 9500 Gilman\\
    Drive, La Jolla, CA 92093}

 \pagerange{\pageref{firstpage}--\pageref{lastpage}} 
 \pubyear{2015}
\maketitle
\label{firstpage} 
\begin{abstract} 
We present FIRE/\texttt{Gizmo} hydrodynamic zoom-in simulations of isolated dark matter halos, two each at the mass of classical dwarf galaxies ($\mvir \simeq 10^{10} \msun$) and ultra-faint galaxies ($\mvir \simeq 10^9 \msun$), and with two feedback implementations.   The resulting central galaxies lie on an extrapolated abundance matching relation from $\mstar \simeq 10^6$ to $10^4 \msun$ without a break.   Every host is filled with subhalos, many of which form stars. Our dwarfs with $\mstar \simeq 10^6~\msun$ each have $1-2$ well-resolved satellites with $\mstar = 3-200 \times 10^3 \msun$. Even our isolated ultra-faint galaxies have star-forming subhalos. If this is representative, dwarf galaxies throughout the universe should commonly host tiny satellite galaxies of their own.  We combine our results with the ELVIS simulations to show that targeting $\sim 50~\kpc$ regions around nearby isolated dwarfs could increase the chances of discovering ultra-faint galaxies by $\sim 35\%$ compared to random halo pointings, and specifically identify the region around the Phoenix dwarf galaxy as a good potential target.
  
The well-resolved ultra-faint galaxies in our simulations ($\mstar \simeq 3 - 30 \times 10^3~\msun$) form within $\mpeak \simeq 0.5 - 3 \times 10^9 \msun$ halos.  Each has a uniformly ancient stellar population ($ > 10~\gyr$) owing to reionization-related quenching.   More massive systems, in contrast, all have late-time star formation. Our results suggest that $\mhalo \simeq 5 \times 10^9 \msun$ is a probable dividing line between halos hosting reionization ``fossils" and those hosting dwarfs that can continue to form stars in isolation after reionization.   
\end{abstract}


\section{Introduction}
\label{sec:intro} 

If \lcdm~is correct, then all dark matter halos hosting galaxies -- from those hosting dwarfs to those hosting giant clusters -- should be filled with substructure \citep{Moore1999, Klypin1999, ZentnerBullock2003}. Dark-matter only simulations over a vast range of particle masses and physical scales show that substructure persists down to the resolution limit of \lcdm~simulations \citep{Diemand2008, Madau2008, Springel2008, Boylan-Kolchin:2009ly, Bolshoi2011}.  The basic expectation is that the mass function of subhalos rises steadily to masses well below the molecular cooling limit of $\mhalo \sim 10^6 \msun$ \citep{Tegmark1997}, with thousands of sites for potentially star-forming satellites.

Observations of our own Milky Way, on the other hand, have revealed the presence of only $\sim 30$ confirmed satellite galaxies \citep{Willman2011, McConnachie:2012, Belokurov2010}, the faintest of which were all discovered by the Sloan Digital Sky Survey (SDSS) \citep{Willman2005b, Willman2005a, Grillmair2006, Sakamoto2006, Irwin2007, Walsh2007, Grillmair2009, Zucker2006a, Zucker2006b, Belokurov2006, Belokurov2007, Belokurov2008, Belokurov2009, Belokurov2010}. However, the SDSS covers only a fraction of the sky and is incomplete to the most distant and faintest satellites of the Milky Way. Any solution to the mismatch between the predicted abundance of subhalos and the observed counts of satellite galaxies around our own Milky Way will likely involve the discovery of ultra-faint satellites in unprobed regions of the sky, at large distances, and with surface brightnesses low enough that they lie just outside current detection limits \citep{Benson2002b, Ricotti2005, Koposov2008, Tollerud2008, Bullock2010}. Over the last several years, searches by PanSTARRS and VST ATLAS have failed to lend critical support to this idea -- finding far fewer dwarf galaxy satellites than expected \citep{Laevens2014, Martin2014}. However, the recent discovery of up to nine new ultra-faint satellites in the southern sky by the Dark Energy Survey \citep[DES,][]{DES2015, Koposov2015}, the detection of three more faint dwarf satellite candidates \citep{Laevens2015, Martin2015, Kim2015}, and ongoing efforts to discover dwarfs at large distances \citep{Tollerud2015} provide an exciting glimpse into the near future and point to a much larger population of as-yet-undiscovered dwarf galaxies in the Local Volume.

Interestingly, most of the newest Milky Way satellite candidates were all discovered near the Large Magellanic Cloud (LMC). While this could be nothing more than a location-based selection effect (the LMC happens to lie next to the region probed by the year one data release of DES), \citet{Deason2015} show that several of the candidates could very well be satellites of the LMC, and their discovery highlights the potential for discovering ultra-faint satellites of other more massive dwarf satellites, or of isolated dwarf galaxies in the Local Group. The scale-free nature of the subhalo mass function in \lcdm~suggests that groups of subhalos should be common \citep{Moore1999, Li2008, Wetzel2015}. Because low-mass halos form earlier, are denser, and fall into smaller hosts before larger ones \citep{NFW1997}, it is likely that satellites of satellites or of lower mass isolated halos may have survived longer than their counterparts that fell directly into the Milky Way \citep{Diemand2008}. This suggests that one way to search for ultra-faint galaxies might be as satellites of known dwarf galaxies. Associations and pairs of satellite galaxies in the Local Group have been observed for some time \citep{Tully2006, DOnghiaLake2008, Ibata2013, Fattahi2013}, and several Milky Way satellites are suspected to host their own satellites \citep{Pace2014, Deason2014}. 

Once a group of satellites falls into the Milky Way, however, tidal forces will eventually disassemble the group and wipe out evidence of coherent structure \citep{Sales2007b, Deason2015}. Isolated massive dwarfs in the Local Group may serve as complementary targets in the hunt for ever-fainter dwarfs (which presumably probe the lowest mass dark matter subhalos). \citet{Bovill2011b} suggest that even already-merged satellites of isolated dwarfs could be detected as ``ghost halos" of ancient stellar populations surrounding their hosts. Despite the obvious challenge of detecting ultra-faints at large distances, isolated dwarfs also make efficient use of telescope pointings, removing some of the chance inherent in any random ultra-faint search without losing any of the search volume. In the near future, large-area, deep-sky surveys such as LSST, DES, PanSTARRS, and SkyMapper will push current detection limits to lower surface brightness and may be able to see ultra-faint satellites of isolated dwarf galaxies if they exist \citep{Ivezic2008, Kaiser2002, Keller2007}. 

Dynamical measurements of most Milky Way dwarf spheroidals (dSphs) show that, despite having luminosities that vary over nearly five orders of magnitude, they have almost the same central densities, comparable to dark matter halos with $\mpeak \sim 3 \times 10^9~\msun$ \citep{Strigari2008, Geha2009, Wolf2010}. This is surprising, not only because this is well above the mass where physical processes are believed to prevent star formation, but because the mass function of potentially star-forming subhalos should be heavily populated by smaller halos. Segue 2 may be the first of these long-searched-for small subhalos to be identified \citep{Kirby2013}, as it appears to have a total dark matter mass $\lesssim 10^8 \msun$.  However, based on its metallicity, \citet{Kirby2013} hypothesize that Segue 2 is a bare remnant of a much larger galaxy that has been severely tidally stripped. The common central density for observed dSphs might signify a low-mass cutoff in galaxy formation. Halos with slightly lower central densities may have been unable to shield themselves from the reionizing background, perhaps due to having a lower baryon fractions than their denser counterparts \citep{Milosavljevic2014}. Alternatively, \citet[][hereafter B10]{Bullock2010} suggest that the apparent common mass scale is a selection effect, and that a large population of unobserved ``stealth galaxies" may reside in halos with masses just below those that host observed dwarfs. These low-mass halos have shallow potential wells, and therefore the galaxies that form within them have larger effective radii and lower surface brightnesses, allowing them to more easily avoid detection \citep{Kaufmann2007, Bullock2010, Bovill2011a, Bovill2011b}. 

The number of these stealth galaxies expected to exist is sensitive to the presence of a low-mass cutoff in galaxy formation, as the lowest mass halos are expected to host the ``puffiest" galaxies in this picture.  The heating of accreted gas or the prevention of gas accretion by the ambient ionizing UV background -- created when the first galaxies formed -- can arrest or prevent star formation in the smallest dark matter halos, suggesting a threshold in halo mass below which all halos remain dark \citep{Efstathiou1992, Somerville2002, Benson2002a, Kravtsov2004, Ricotti2005, Moore2006, Strigari2007, Simon2007, Madau2008}. It is possible, however, that some very low mass halos were able to form some Population II stars \textit{before} reionization \citep{Bullock2000}. If star formation ceased after this time, it would cause them to have uniformly ancient stellar populations \citep{Ricotti2005} with extremely low metallicity \citep{Bovill2009, Ricotti2010}.  Remarkably, this appears to be the case with the known ultra faint dwarfs of the Milky Way \citep{Brown2014}. 

Several authors have shown that photoheating prevents gas from condensing in halos with masses that lie just below or at the common halo mass scale for Milky Way satellites \citep{Okamoto2009, Nickerson2011, Sawala2014, Shen2014}.  However, the specific timing of the onset and end of reionization, the spectrum of the background radiation, the ability of gas to temporarily shield itself from these high energy photons, and particularly the mass resolution, can have a large effect on the number and minimum halo mass of galaxies in any simulation \citep{Efstathiou1992, Dijkstra2004, Hoeft2006, Onorbe2015}. Moreover, prescriptions for how stars and supernovae return energy back to the interstellar medium (ISM) will affect the formation history of small galaxies in the presence of an ionizing background.  

In what follows, we present a series of simulations of galaxy formation within small dark matter halos run with the \texttt{Gizmo} code \citep{Hopkins2014b} in ``PSPH-mode" \citep{Hopkins2013}. \texttt{Gizmo} implements the Feedback in Realistic Environments (FIRE)  \citep{Hopkins2014a} feedback scheme for converting gas into stars and capturing the energy fed back from those stars into the surrounding medium. Using these simulations, we find that the subhalos of the halos that surround $\mstar \sim 10^6~\msun$ dwarf galaxies should form stars fairly abundantly, and produce potentially observable ultra-faint satellite galaxies of known classical dwarfs in the Local Group. Further, the ultra-faint dwarfs in our simulations with $\mstar = 3-30 \times 10^3~\msun$ that form in isolation or as satellites within subhalos of $\mpeak \simeq 0.5-3 \times 10^9 \msun$ all have completely ancient stellar populations, as is seen for the known population of the Milky Way. They also have low-surface brightnesses, as expected for ``stealth galaxies" forming in halos of these low masses, but may be within the detection capabilities of future surveys.  

This paper is organized as follows: in Section \ref{sec:sims}, we give the details of our simulations, including halo finding and merger-tree analysis. Section \ref{sec:results} outlines our main results, including the $\rm \mstar - M_{halo}$ relation for our galaxies and their star formation histories. In Section \ref{sec:det}, we use the Exploring the Local Volume in Simulations (ELVIS) suite of dark-matter-only Local Group simulations \citep{Garrison-Kimmel2014a} to determine how common these objects should be and the likelihood of their detection. We compare to several recent works in Section \ref{sec:compare}, and summarize our results and conclude in Section \ref{sec:dis}.

\begin{figure*}
\centering
	\includegraphics[scale=0.55]{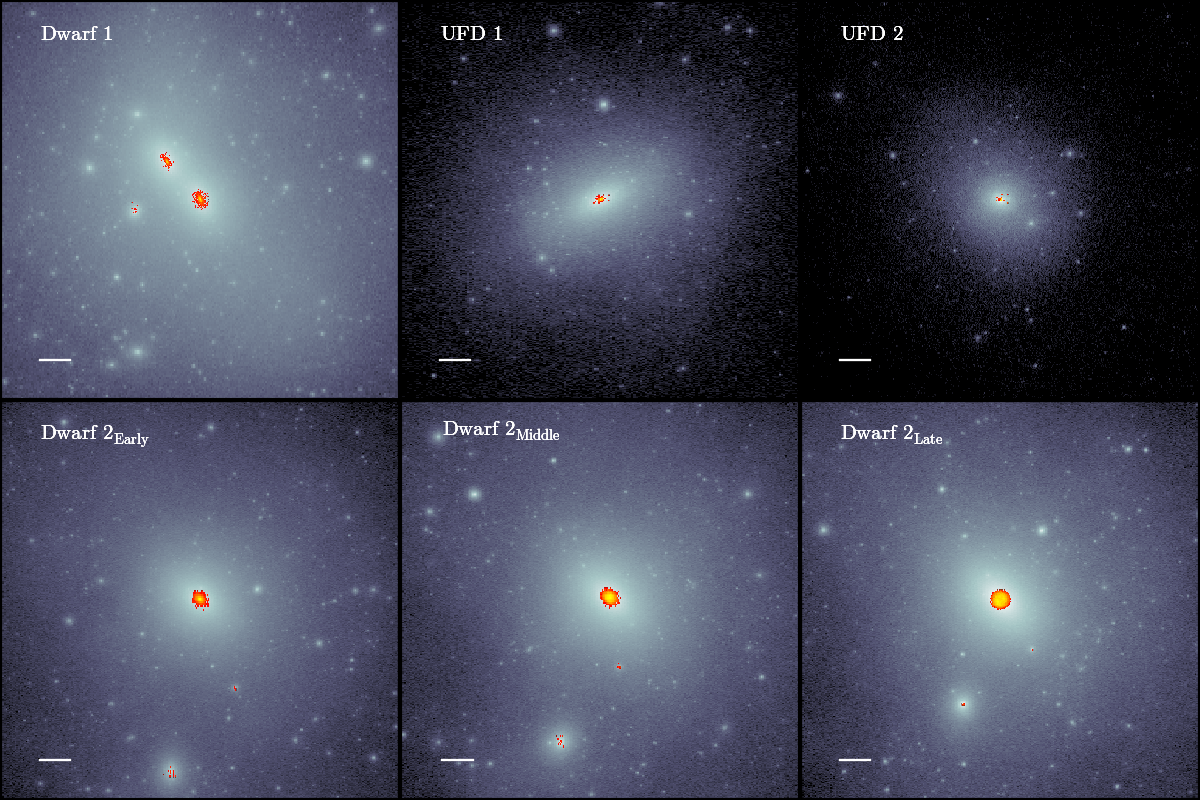}
 	\caption{Projected ($x-y$ plane) visualizations for all of the runs presented in this work. From left to right, the top three panels visualize $\done$, $\ufdone$ and $\ufdtwo$, while $\dearly$, $\dmid$ and $\dlate$ are shown in the lower panels. The dark matter distribution is shown in greyscale, while the stellar density is pictured in color for all subhalos with $\mstar > 3 \times 10^3~\msun$ (in at least one run in the case of $\dtwo$; see text for details). $\mstar$ is calculated as the stellar mass within the inner $3~(1.5)~\kpc$ of the central (satellite) galaxy. The white line in the lower left of each panel represents $5~\kpc$.}
 \label{fig:DM_hists}
\end{figure*}

\section{Simulations}
\label{sec:sims}

We present six cosmological zoom-in simulations of four isolated dwarf galaxy halos using two implementations of subgrid physics. The four host halos are $\done$ and $\dtwo$, which have $z=0$ virial masses of $\mvir \simeq 10^{10}~\msun$; and $\ufdone$ and $\ufdtwo$, which have $\mvir \simeq 10^{9}~\msun$ (see Table 1 for precise numbers).  The $\dtwo$ halo was simulated three times -- two runs varied the specific subgrid feedback implementation, while the third used a different choice of gravitational softening (see below), resulting in changes in the star formation histories of the main galaxies (see \citealt{Onorbe2015} for details).  We refer to these three cases as $\dearly$ (run with the fiducial parameters -- the same parameters as $\done$, $\ufdone$, and $\ufdtwo$), $\dmid$, and $\dlate$.  The subscripts refer to when the halos formed their stars. Every run has a dark matter force softening of $25$ to $35$ pc and a dark matter particle mass of $m_p^{\rm dm} = 1.26 \times 10^3 \msun$ except $\ufdtwo$, which uses $m_p^{\rm dm} = 2.46 \times 10^{3} \msun$.  The $\ufdone$ and three $\dtwo$ simulations are identical to the ``Ultrafaint" and Dwarf runs presented by \citet{Onorbe2015}.~\footnote{Note that $\dearly$ and $\ufdone$ were also previously presented in \citet{Hopkins2014a} as $\rm m9$ and $\rm m10$ respectively.}

The initial conditions were generated using \texttt{MUSIC} \citep{Hahn2011}, and selected from a cosmological box run at low resolution to $z=0$. $\done$, $\dtwo$, and $\ufdone$ were all selected from $5~\hmpc$ boxes to have typical values of spin parameter $\lambda$, concentration, and formation time for their mass range, and also to have small Lagrangian volumes \citep{Onorbe2014}. $\ufdtwo$ was selected from a $25~\hmpc$ box and required to have no other halos of $50\%$ or more of its mass within $4~\rvir$ at $z = 0$ and a small Lagrangian volume. All dwarfs in this work are isolated. Once the halos are identified, the particles are traced back in time and an enclosing Lagrangian volume is chosen and re-simulated at higher resolution with dark matter and gas, buffered by dark matter-only regions of increasing particle mass. This process is done according to the zoom-in techniques outlined by \citet{Katz1993, Onorbe2014} and with the goal of minimizing low resolution particles within the halos of interest. All simulations begin at a redshift of $z=125$ and have $0\%$ contamination from low resolution particles within a distance of $1.6~\rvir$ at all redshifts.

All simulations were run using the fully conservative cosmological hydrodynamic code \texttt{Gizmo} \citep{Hopkins2014b} in ``PSPH-mode". This code adopts the Lagrangian ``pressure-entropy" formulation of the smooth particle hydrodynamic (SPH) equations, ameliorating previous difficulties SPH codes had in modeling multiphase fluids \citep{Agertz:2007}. 

For feedback and star formation we use the Feedback in Realistic Environments (FIRE) scheme \citep{Hopkins2014a}.  FIRE tracks momentum imparted locally from stellar radiation pressure, radiation pressure on larger scales via the light that escapes star-forming regions, HII photoionization heating, supernovae (Type I and II) heating, momentum and mass loss and stellar winds from O-type and AGB stars. Each star particle has an age determined by its formation time, an initial mass function (IMF) taken from \citet{Kroupa2002}, and a metallicity inherited from its parent gas particle. The star particle then loses mass and creates metals according to the \texttt{STARBURST99} stellar population synthesis model \citep{Leitherer1999} and the input IMF. Our simulations use an ambient ionizing UV background from \citet{Faucher-Guigere2009}, which starts at $z = 10.65$ and completes reionization by $z \simeq 6$.

The FIRE implementation has as few free parameters as possible, with star formation in the multiphase ISM naturally self-regulating. Giant molecular clouds (GMCs) form, and then are subsequently heated and disrupted by the stars that form within the self-gravitating, molecular gas. In order to ensure that the gas within the GMCs reaches a density high enough to form stars ($n > 100~\cmcb$), the code requires extremely high mass and spatial resolution. Every run uses a gas particle mass of $m^{\rm gas}_{p}  = 255~\msun$ except for $\ufdtwo$, which uses $m^{\rm gas}_{p} = 499~\msun$. The gas force resolution varies from $\rm \epsilon^{min}_{gas} = 1.0 - 2.8~\pc$ in all runs except one version of $\dtwo$ (see below).

The $\dearly$ run uses the fiducial subgrid parameters.  In $\dmid$ and $\dlate$, mass, momentum, and energy are deposited to particles within the SPH kernel according to a mass-weighting scheme, rather than volume-weighting as in the fiducial runs. In addition to this change, $\dmid$ was run to test equal softening lengths for the dark matter and the gas (25 pc). All three runs of $\dtwo$ end up with stellar masses in the central galaxies that differ by only $\sim 25\%$, suggesting that the major differences between the runs are more a result of stochasticity than the changes to the feedback implementation and softening length (see Section \ref{sec:sample}).

We use the Amiga Halo Finder \citep[\texttt{AHF},][]{Knebe2009} to identify gravitationally bound dark matter, stellar and gas particles in each snapshot of the simulation. Because pure particle-matching algorithms generally underperform relative to more sophisticated methods in terms of detection and removal of spurious interlopers between snapshots \citep{Srisawat2013}, we track particle overdensities between simulation snapshots using the \texttt{consistent-trees} software \citep{Behroozi2013c}. This is particularly important given our focus on subhalos. Because the current stable version of \texttt{consistent-trees} -- which makes use of the \texttt{ROCKSTAR} \citep{Behroozi2013} halo finder -- does not track baryonic particles, we build a pipeline between \texttt{AHF} and \texttt{ROCKSTAR} that allows for a seamless transition between the \texttt{AHF} halo catalogs and the merger trees. This pipeline uses the halo phase space information combined with the dark matter particle IDs to match each halo in the catalogue to a \texttt{consistent-trees} merger tree. \texttt{ROCKSTAR} and \texttt{consistent-trees} were used to determine the peak virial mass and infall times for subhalos. All other halo, galaxy and stellar quantities were obtained from the \texttt{AHF} halo catalogs or the raw particle files.

 All runs were initialized and simulated assuming the WMAP-7 cosmology $\sigma_8 = 0.801$, $\Omega_{\Lambda} = 0.734$, $\Omega_m = 0.266$, $\Omega_b = 0.0449$, $n_s = 0.963$ and $h = 0.71$ \citep{Komatsu2011}. Further detail on the FIRE/\texttt{Gizmo} code can be found in \citet{Hopkins2014a, Hopkins2014b}.

\begin{table*}
\centering 
\begin{tabular}{| l  | c | c c c | c  | c | } 
\hline
   & $\done$ & $\dearly$ & $\dmid$  & $\dlate$ & $\ufdone$   & $\ufdtwo$   \\
\hline\hline
{\hspace{.2in}} Central & & & & & & \\
\hline
$\mvir \, (10^9 \msun)$  & 11.1 & 7.8 & 7.7 & 7.6 & 2.5  & 1.1 \\
$\mstar \, (10^3 \msun)$  & 620 & 2200 & 2700 & 2800 & 22 & 8.5  \\
\hline
{\hspace{.1in}} 1st Satellite & & & & & &  \\
\hline
$\mpeak \, (10^9 \msun)$  & 4.7 & 0.70 & 0.68  & 0.64 &  0.03 & 0.02 \\
$\mstar \, (10^3 \msun)$  & 220 & 4.2 & 5.4 & 2.9 & (0.9) & (0.7) \\
\hline 
{\hspace{.1in}} 2nd Satellite & & & & &  & \\
\hline
$\mpeak \, (10^9 \msun)$  & 0.51 &  0.04 & 0.04 & 0.04 &  0.02 & 0.007 \\
$\mstar \, (10^3 \msun)$  & 5.0 & (0.7) & (3.5) & (0.6) & (0.6)  & (0.4) \\
\hline 
\end{tabular} 
\label{tab:tablebig} 
\caption{Halo and stellar masses of all central galaxies and their 1st and 2nd Satellites from all six simulations. For all simulations except $\dearly$ and $\dlate$, the 1st and 2nd Satellites refer to the first and second most massive satellites in stellar mass. The 2nd Satellites for $\dearly$ and $\dlate$ are selected because their counterpart within $\dmid$ has $\mstar > 3 \times 10^3~\msun$. Both satellites in $\done$ and the 1st Satellites in all $\dtwo$ runs are ``massive satellites" (see text). The stellar masses of all other satellites are shown in parentheses.}
\end{table*}

\subsection{Sample Details}
\label{sec:sample}

Figure \ref{fig:DM_hists} shows the dark matter distributions of all six runs in greyscale with the stellar mass density overlaid in color. The physical scales of the panels are identical:  the white line in the lower left corner of each visualization indicates $5~\kpc$. 

For our detailed analysis, we focus on objects with $\mstar > 3 \times 10^3~\msun$ in order to avoid effects from spurious star formation at our resolution limit. The gas density required to form stars in our simulations is quite high and can only occur in dense, molecular, self-gravitating regions. This onerous minimal criteria suggests that the formation of this minimum stellar mass represents a physical star formation episode. \footnote{All galaxies that meet our minimum $\mstar$ requirement have at least 16 star particles and had $\gtrsim 3 \times 10^4$ dark matter particles at peak mass. All but one had $\gtrsim 2 \times 10^5$ dark matter particles at peak mass, corresponding to typical initial baryonic particle numbers of $\sim 3-4.5 \times 10^5$ before feedback dispelled most of these.} Furthermore, we find that below a subhalo mass of $\mpeak \sim 5 \times 10^8~\msun$ our results become far less stable against the stochasticity in star formation (see below). We therefore consider only as robust those objects that form within halos that have a minimum dark matter particle number $N^{\rm dm}_{p} \geq 2 \times 10^5$ and refer to satellites that meet both this requirement and the stellar mass cut as ``massive satellites". 

$\done$ has two satellite galaxies that meet both of these requirements. Only one subhalo -- the most massive in each of the runs of $\dtwo$ -- forms a ``massive satellite". In $\dmid$ there is a second satellite that meets the stellar mass but not the dark matter particle number cut. Because all three runs of $\dtwo$ use identical initial conditions, each subhalo in $\dmid$ has a corresponding subhalo in the other runs that shares a majority of the dark matter particles. The subhalos corresponding to this second satellite host galaxies below the stellar mass cut in $\dearly$ and $\dlate$. In Figure \ref{fig:DM_hists}, stars are shown in this particular subhalo for all three runs of $\dtwo$. Importantly, although the three subhalos that host the second satellites of $\dtwo$ are effectively identical, they produce galaxies that differ by a factor of $\sim 6$ in stellar mass due to only minor changes in parameters between the runs. This motivates the fairly large dark matter particle number we require to consider a system well resolved.

$\mhalo$ and $\mstar$ for the central galaxy (Central), most-massive satellite (1st Satellite), and second-most-massive satellite (2nd Satellite) \footnote{In all cases except $\dearly$ and $\dlate$, the 2nd Satellite is the second most massive satellite in stellar mass. In these two runs only, the 2nd Satellites are selected as the satellites that correspond to the high stellar mass outlier in run $\dmid$. Only the 1st Satellites in $\dtwo$ runs qualify as ``massive satellites".} in all six runs can be found in Table 1. In the Tables, Figures and throughout the text, $\mhalo$ is $\mvir$ for centrals and $\mpeak$ for satellites. Values in parentheses indicate satellite systems that we regard as too poorly resolved to trust for detailed analysis.  They are included here for completeness.  The 2nd Satellites of all $\dtwo$ runs fall into this category, as well as all satellites of the isolated ultra-faints $\ufdone$ and $\ufdtwo$.

In Figure \ref{fig:DM_hists}, the dark matter distribution is shown for all dark matter particles located within $65\%$ of the virial radius of the halo listed in each panel. Star particles are only shown if they reside within any halo that currently has $ > 3 \times 10^3~\msun$ in stars in the inner $3~(1.5)~\kpc$ of a central (satellite) galaxy or are in a corresponding halo in the other two runs of $\dtwo$. Many of the other subhalos also contain star particles (see Figure 2), but they are not shown so that we may focus on systems that are reasonably well resolved.  The radial extent of the galaxies was chosen by inspecting their stellar mass profiles (each central galaxy also has a {\em very} diffuse stellar halo that we aim to avoid). The galaxy stellar masses are all computed using these radial limits.

\begin{figure*}
\hspace{-0.75cm}
 \includegraphics[scale=0.18]{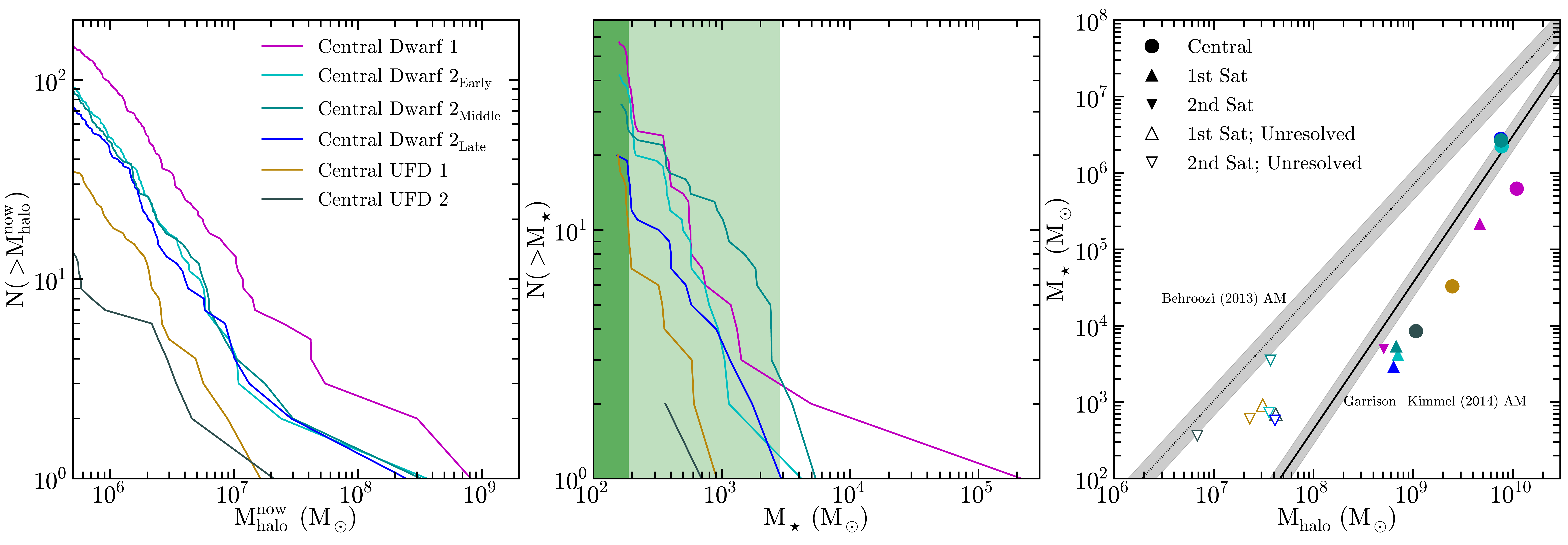}
 \caption{Left: Subhalo mass function for all six runs. $M^{z=0}_{\rm halo}$ is the present day mass for all subhalos. The left edge of the panel corresponds to $\sim 400$ bound dark matter particles. Note: if these halos simply contained the universal fraction ($f_{\rm b} = 0.169$) of baryons, this would correspond to $\gtrsim 350$ baryonic particles in these halos before they were forcefully removed by feedback. Middle: Stellar mass function for all satellites that form at least one star particle. The average mass of a single star particle is shown as a dark green band. Masses we regard as poorly-resolved ($\mstar < 3 \times 10^3~\msun$) are highlighted by the light green region. Right: $\mstar - \mhalo$ for all Centrals (filled circles), 1st and 2nd satellites (upward and downward triangles respectively). With surprisingly little scatter, our isolated dwarfs and their six well-resolved satellites (filled triangles) sit just offset from the extrapolated abundance matching relation from \citet{Garrison-Kimmel2014a} (solid black line). Both this and the abundance-matching relationship from \citet{Behroozi2013f} (dotted black line) are shown with $0.2$ dex in scatter. Open symbols correspond to poorly resolved galaxies: $\mstar < 3 \times 10^3~\msun$ and/or $N^{\rm dm}_{p}< 2 \times 10^5$.
 \label{fig:3panel}
}
\end{figure*}
 
\section{Results}
\label{sec:results}

In all six runs, the dark matter halos hosting the central galaxies contain significant substructure. The leftmost panel of Figure \ref{fig:3panel} shows the current subhalo mass function for all six runs. We plot all subhalos with $M^{z=0}_{\rm halo} > 5 \times 10^5~\msun$, which corresponds to $N^{\rm dm}_{p} \gtrsim 400$. $\done$, with the highest virial mass, has both the most and the most massive subhalos. The subhalo mass functions for the three runs of $\dtwo$ lie predictably on top of each other. As expected, the two isolated ``ultra-faints" ($\mvir < 3 \times 10^9~\msun$) form far fewer subhalos (at fixed mass) than their more massive counterparts. 

The satellite stellar mass functions are shown in the middle panel of Figure \ref{fig:3panel}. They are intriguingly steep, rising to as many as fifty tiny satellites in the case of $\done$.  However, the majority of these star-forming satellites form only a single star particle, and this potentially exciting behavior will need to be confirmed at much higher resolution.  In order to make this clear, the dark green shaded region shows the average stellar mass of a single star particle, while the light green shaded region shows the mass range for all other objects that form with $\mstar < 3 \times 10^3~\msun$ (fewer than 16 star particles). In all six simulations, a total of $187$ subhalos form at least one star particle, but a mere six of them form galaxies with $\mstar > 3 \times 10^3~\msun$ (and only five of these also meet the $N^{\rm dm}_{p}$ requirement). It is these, the most massive satellites, that will be the focus of our analysis (along with the centrals themselves).

The $\mstar - \mhalo$ relation for the five resulting ``massive satellites" (filled triangles), their less-well-resolved counterparts (open triangles), and all central galaxies (filled circles), is shown in the right panel of Figure \ref{fig:3panel}. Plotted are the six central galaxies along with the 1st and 2nd Satellites of each host (detailed in Table 1) for each run. Although 18 symbols are shown, we only regard the 11 filled points as reasonably well-resolved. All of the well-resolved systems in this work lie only slightly offset from the extrapolated abundance matching relationship from \citet{Garrison-Kimmel2014a}, which has been constrained down to a stellar mass of $\sim 10^5 \msun$. This abundance matching relation, obtained by updating the \citet{Behroozi2013f} relation with a faint-end slope derived from the measurements of the GAMA survey \citep{Baldry2012}, accurately reproduces the stellar mass functions of the Milky Way and Andromeda satellites, as well as the stellar mass function of the Local Field, when applied to the ELVIS simulations \citep{Garrison-Kimmel2014a}. 

We do not find the ``bending" of the $\rm \mstar - M_{halo}$ relation reported by \citet{Sawala2015} at $\mhalo \simeq 5 \times 10^9~\msun$. Instead, the relation maintains a fairly steady power-law down to $\mhalo \simeq 10^9 \msun$.  We do see a qualitatively similar break to that seen by \citet{Sawala2015} but at a significantly lower halo mass of $\mhalo \simeq 5 \times 10^8~\msun$.  Below this point it is clear that we are witnessing a large amount of stochasticity, the onset of which corresponds to the scale of extremely faint galaxies ($\mstar \sim 10^3~\msun$), as opposed to the results of \citet{Sawala2015}, where stochasticity sets in near the masses of classical dSphs ($\mstar = 10^5~\msun$). However, it is in this regime where our resolution is severely limited -- these galaxies have $\lesssim 15$ star particles. At this time we are unable to determine whether the apparent break and associated stochasticity in our relation at small mass are physical or simply artifacts of resolution, although it is intriguing that the break witnessed by \citet{Sawala2015} occurs at the mass of $\sim 10$ of their baryonic particles as well. Future work at higher resolution will allow us to explore this question more fully. 

The most massive satellite of $\done$ represents an interesting statistical rarity (see Section \ref{sec:elvis}) in that the mass of the subhalo hosting this satellite is just over $\rm 40\%$ of the mass of its host, and the satellite itself has $1/3$ of the stellar mass of its central galaxy.  $\done$ is clearly undergoing a major merger (see Figure \ref{fig:DM_hists}) and the associated large satellite has properties that are distinct from the other satellites described in this work. Most notably, its stellar mass is two orders of magnitude greater than the other ``massive satellites", and it’s stellar and dark matter masses are both larger than those of the central hosts of $\ufdone$ and $\ufdtwo$. For identification only (with no relation to the structural properties of the satellite), we refer when necessary to this most massive satellite as ``dSph", and all other satellites as ``ultra-faint" satellites. The broader term ``ultra-faints" is reserved for all galaxies, centrals and satellites, that have $\mstar = 3-30 \times 10^3~\msun$ in our simulations.

\subsection{Star Formation Histories}
\label{sec:sfh}

\begin{figure}
\includegraphics[scale=0.25]{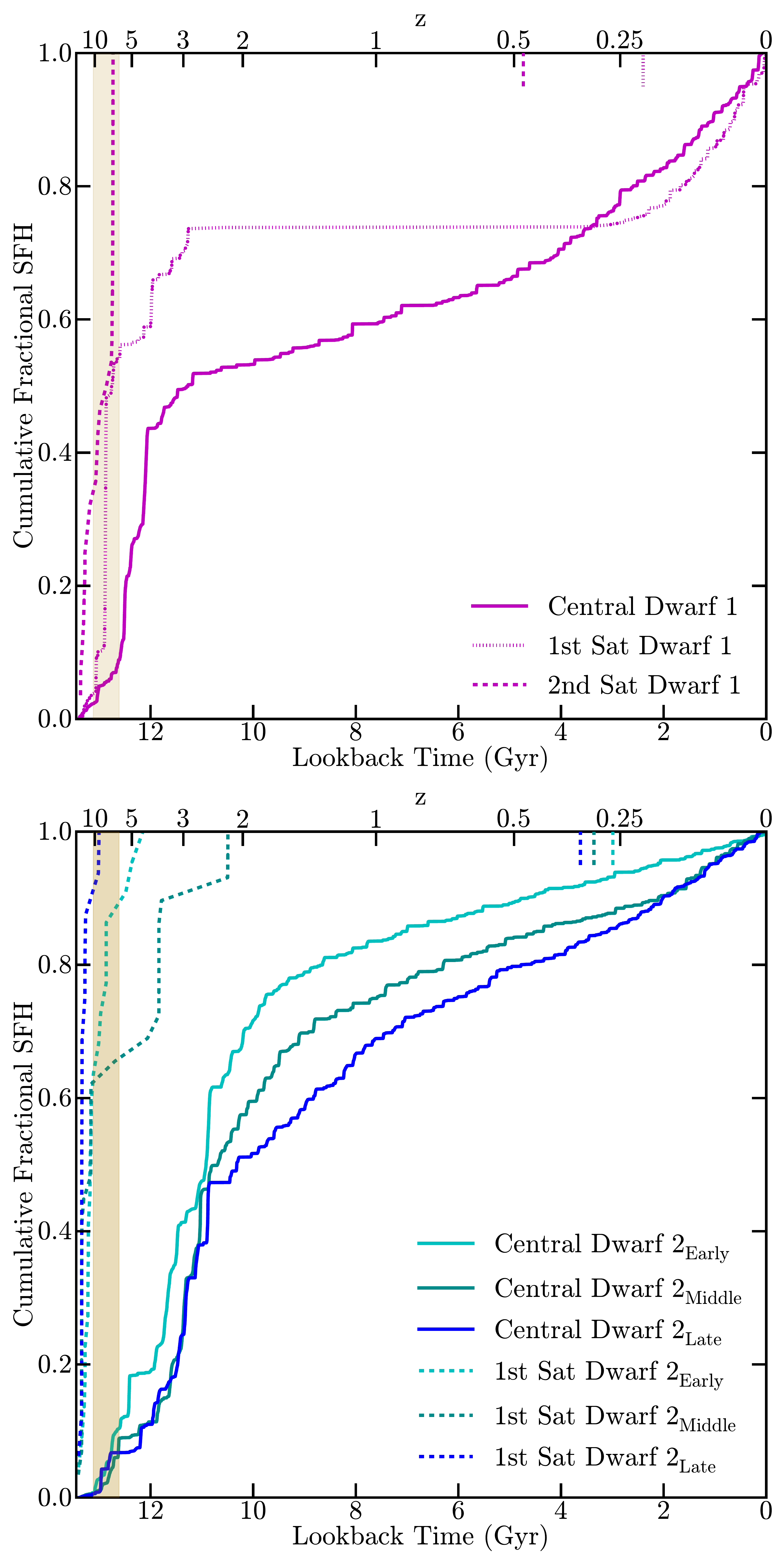}
 	\caption{Cumulative fractional star formation histories (SFH) of the Centrals (solid) and well-resolved satellites (dashed / dotted) in the two isolated Dwarf simulations ($\mvir \sim 10^{10}~\msun$). Short vertical lines along the top axes indicate infall times of the satellites.  Top: $\rm Dwarf \, 1$ Central (solid) along with its two satellites.  The massive ``dSph" satellite is shown by the dotted line and it displays a late-time burst in conjunction with its recent infall. The ``ultra-faint" satellite (dashed) is quenched early, during reionization (shaded band), well before it is accreted.   Bottom: $\rm Dwarf \, 2$ centrals (solid) and the most massive (ultra-faint) satellites (dashed) in each run. 
 \label{fig:SFH_h12}
 }
\end{figure}

The top panel of Figure \ref{fig:SFH_h12} shows the cumulative fractional star formation histories (SFH) for $\done$ (solid line) and its two ``massive satellites".  The SFH of its large, ``dSph" satellite ($\mstar^{z=0} \simeq 2 \times 10^5 \msun$) is shown by the dotted line. Its second, ``ultra-faint" satellite ($\mstar^{z=0} \simeq 5 \times 10^3 \msun$) is shown by the dashed line.  The shaded band corresponds to the epoch of reionization in our simulations.    The bottom panel presents similar histories for the three $\dtwo$ runs along with the single most massive ``ultra-faint"  satellite ($\mstar^{z=0}  \simeq 3-5 \times 10^3 \msun$) that forms in each run (dashed lines).  In both panels, the short vertical lines at the top mark the first virial crossing of the satellites.

The massive ``dSph" satellite of $\done$ demonstrates a particularly interesting SFH.  It stops forming stars shortly after the epoch of reionization for almost $\rm 8~\gyr$ (from $z \sim 3$ to $z \sim 0.25$).  However, it manages to keep some of its gas on hand in a state that is just below the threshold for star formation.  Upon infall into the virial radius of $\done$ at a lookback time of $\rm \sim 2~\gyr$, followed by pericentric passage $\sim 500~\myr$ later, its ISM is compressed enough to create a new burst of stars, all of which are concentrated at the center of the satellite and were formed in situ. The central galaxy of $\done$ itself also appears to experience a related (albeit mild) burst. This  type of bimodal star formation history was suggested by \citet{Ricotti2010} as a signature of massive reionization ``fossils".  Our ``dSph" satellite has a slightly higher $\vmax$ than the range given in that work, and its rebirth of star-formation was triggered by a merger rather than by a change in its halo concentration (as originally suggested by \citealt{Ricotti2010}); it is nevertheless similar in its qualitative nature. 

In our simulations, halos with $M \simeq 10^{10} \msun$ appear to robustly maintain star formation to $z=0$. This massive satellite at $\mpeak \simeq 5 \times 10^9 \msun$ seems to mark a transition point where reionization begins to favor uniformly ancient star formation.   The influence of its impending merger was enough to trigger a rebirth.   The ``ultra-faint" satellites shown in Figure \ref{fig:SFH_h12} all form in halos smaller than this, with $\mpeak \lesssim 2 \times 10^{9} \msun$. They all form their stars entirely before $z \sim 2$ and are quenched over $5~\gyr$ before their infall onto the larger hosts, indicating that their star formation is shut down by reionization rather than environmental processes. Importantly, these subsequent mergers do not trigger fresh star formation, as was seen in the more massive satellite. These small halos are either quenched during reionization or shortly thereafter, running out of fuel after the fresh gas supply was shut off by the ambient UV field.  

\begin{figure}
\includegraphics[scale=0.25]{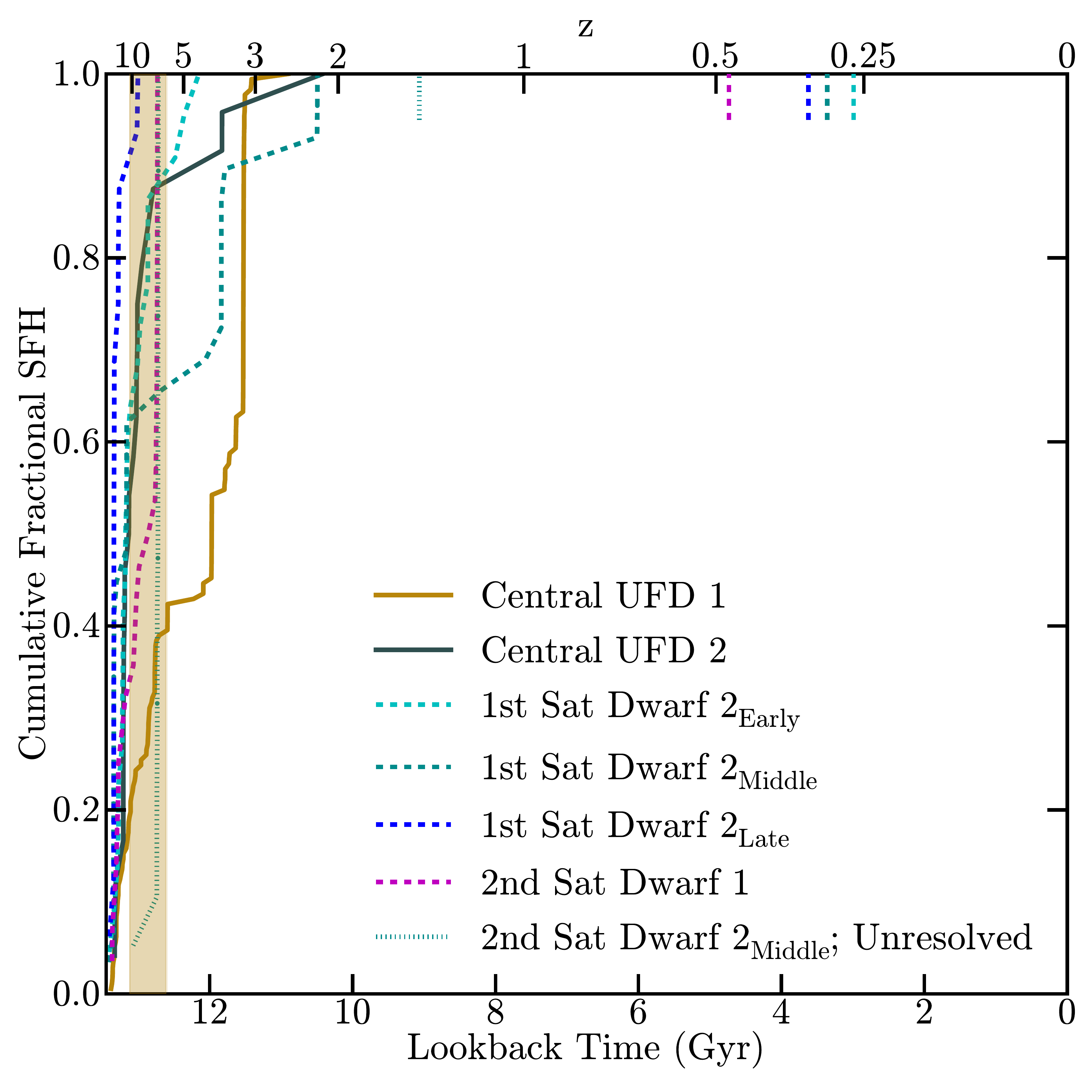}
 	\caption{Ancient ``ultra-faints".  Shown are the cumulative fractional star formation histories (SFH) for our isolated (solid) and resolved satellite (dashed) ``ultra-faint" dwarfs ($\mstar^{z=0}  = 3 - 30 \times 10^3 \msun$). The SFH of the unresolved outlier that passes the stellar mass cut (see Section \ref{sec:sample} for details) is also shown (dotted line). All seven of these systems have ancient stellar populations, similar to those observed for the ultra-faint galaxies of the Milky Way.
 \label{fig:SFH_ufds}
 }
\end{figure}

This tendency for our ultra-faint galaxies  in $\mpeak \lesssim 2 \times 10^{9} \msun$ halos to quench early is illustrated more clearly in Figure \ref{fig:SFH_ufds}, where we have plotted the cumulative fractional star formation histories for all seven ``ultra-faints" formed in our simulations, e.g. all ``ultra-faint" satellites from Figure \ref{fig:SFH_h12} (dashed lines), along with $\ufdone$ and $\ufdtwo$ (solid lines). We also include the unresolved 2nd Satellite of $\dmid$ (dotted line) due to it's high stellar mass and to illustrate its similar SFH. From Figure \ref{fig:SFH_ufds} it is clear that, in every case, reionization plays a significant role in the shutting down of star formation in our ``ultra-faints".  Not only were $100\%$ of all stars in all seven of these objects formed before $z=2$, four of the objects form over $\rm 90\%$ of their stars before the completion of reionization at $z=6$. 

\section{Detecting Satellites of Dwarfs}
\label{sec:det}
\subsection{How Common are Satellites of Dwarfs?}
\label{sec:elvis}

It is beyond the scope of this paper to investigate a statistically significant sample of hydrodynamic simulations of $\mvir \simeq 10^{10}~\msun$ dwarfs in order to estimate the frequency with which they will host satellites above a given mass. However, if we assume an $\mstar$ - $\mpeak$ relation similar to that presented in Figure 2, we can make an estimate using  dark-matter-only simulations.   We do so using the Exploring the Local Volume in Simulations (ELVIS) suite of collisionless zoom-in simulations of Local Group-like environments \citep{Garrison-Kimmel2014a}.   

We select all isolated ($d_{\rm halo} > 2~\rvir^{\rm halo}$ for all more massive halos) dwarf-size dark matter halos ($35~\kms < \vmax < 45~\kms$) in the 12 Local Group-like pairs and the 24 isolated ELVIS simulations, and determine the fraction of those hosts that have $N$ subhalos with $\mpeak \geq 5 \times 10^8~\msun$.  This is the halo mass that corresponds to $\mstar > 3000 \msun$ according to Figure 2. \footnote{In all six of our runs, every subhalo with $\mpeak \geq 5 \times 10^8~\msun$  forms a well-resolved satellite. Although we focus here on satellites of isolated dwarfs, we note that this criteria would suggest the existence of $\sim 100$ undetected ultra-faint ``stealth galaxies" within $400~\kpc$ of the Milky Way.}  We similarly compute the fraction of dwarf halos that have a very large subhalo with $\mpeak \geq 4.5 \times 10^9~\msun$, set by the most massive satellite of $\done~(\mstar \gtrsim 2 \times 10^5 \msun$). 

\begin{figure}
 \centering
 \includegraphics[scale=0.25]{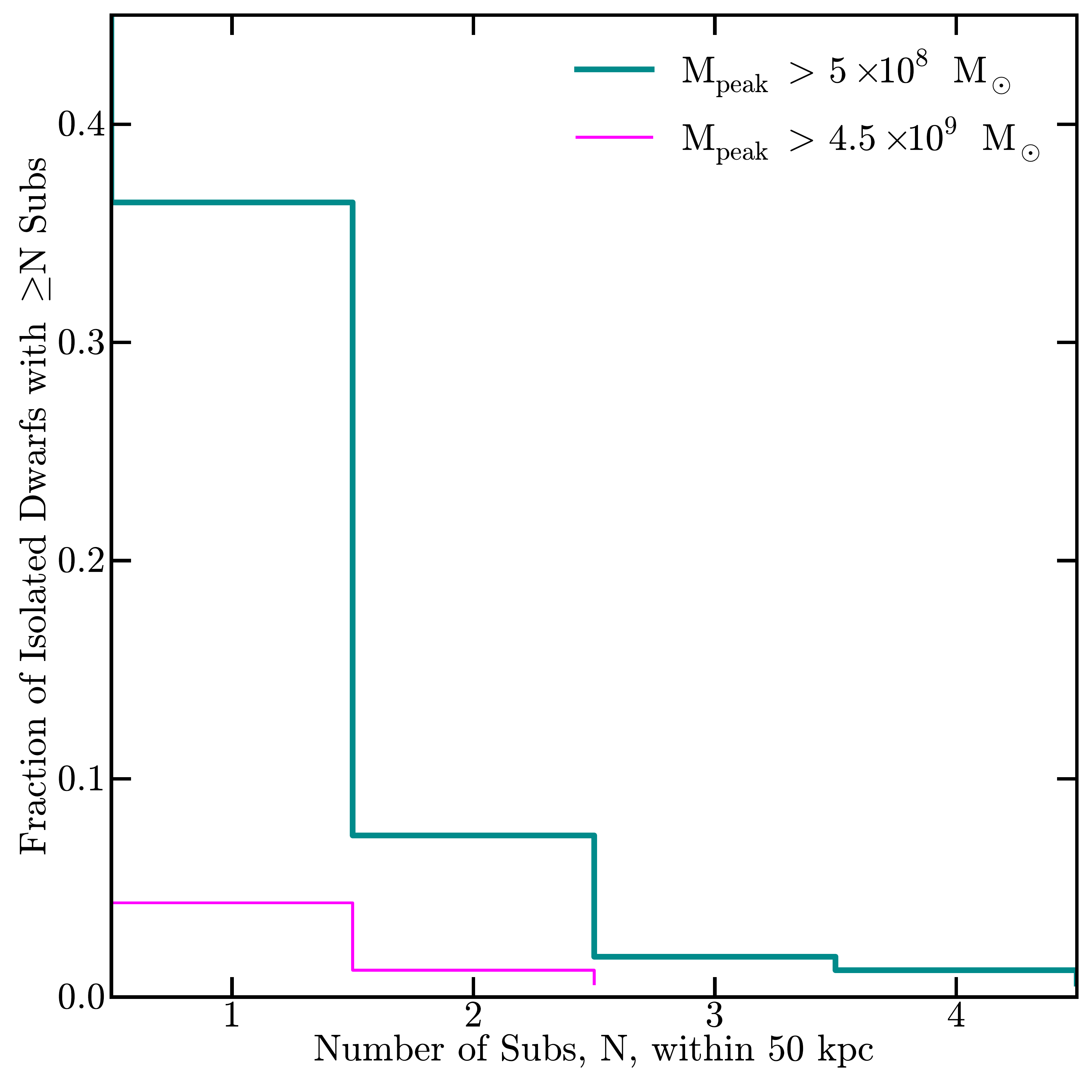}
 \caption{From the dark-matter only ELVIS suite of Local Group simulations, the fraction of isolated dwarfs in the $\vmax$ range $35-45~\kms$ with N or more subhalos with $\mpeak > 5 \times 10^8~\msun$ (thick line) and $\rm 4.5 \times 10^9~\msun$ (thin line) that are found within $50~\kpc$ of their host galaxy. In all of our hydrodynamic runs, all subhalos with $\mpeak~>~5 \times 10^8~\msun$ have formed a galaxy with $\mstar > 3 \times 10^3~\msun$, while the most massive satellite, ``dSph", forms in a subhalo with $\mpeak \simeq 4.5 \times 10^9~\msun$. According to these dark-matter only simulations, $\rm \sim 35\%$ of isolated dwarfs should have a ``massive satellite" within their virial radii, but only $\rm \leq 5\%$ should host a satellite as massive as ``dSph".
 \label{fig:ngtN}
}
\end{figure}

Figure \ref{fig:ngtN} shows the probability for an isolated dwarf halo in ELVIS to have $\rm \geq N$ satellites of at least these two peak virial masses within a projected distance of $50~\kpc$ (the typical virial radius for our dwarfs) as a function of $\rm N$. According to Figure \ref{fig:ngtN}, isolated halos with $\mvir \sim 10^{10}~\msun$ will have one or more subhalos that could host $\mstar \gtrsim 3000 \msun$ satellites about $\rm 35\%$ of the time. The likelihood that an isolated dwarf in the same mass range hosts a satellite as massive $\mstar \simeq 2 \times 10^5 \msun$, however, is just under $5\%$.

Using both SDSS and the semi-analytic models of \citet{Guo:2011} applied to the \msii~simulation \citep{Boylan-Kolchin:2009ly}, \citet{Sales2013} show that the probability for a central galaxy to host a satellite a given fraction of its stellar mass, $\mstar^{\rm sat} / \mstar^{\rm cen} $, decreases as a strong function of $\mstar^{\rm cen}$ until $\mstar^{\rm cen} = 10^{10}~\msun$, becoming independent of $\mstar^{\rm cen}$ for centrals below the stellar mass of the Milky Way. Although the stellar masses of the central galaxies in our simulations lie far below their stellar mass range ($10^{7.5} \leq \mstar / \msun \leq 10^{11}$), this decoupling of the probability from $\mstar^{\rm cen}$ for satellites of dwarf galaxies allows us to make a comparison with the probabilities they find. Their Figure 2 suggests that for all central galaxies with $\mstar^{\rm cen} \leq 10^{10}~\msun$, there is a $\sim 40\% - 50\%$ chance that they host a satellite with stellar mass $\mstar^{\rm sat} / \mstar^{\rm cen} \sim 0.5\%$ -- similar to our ``ultra-faints". This is roughly consistent with the $\sim \rm 35\%$ probability we find. They also find that the likelihood for a satellite as massive in proportion to its central as ``dSph" ($\mstar^{\rm sat} / \mstar^{\rm cen} \sim 1/3$), is vanishingly small. Again, according to their Figure 2 and assuming that the lack of dependence of this probability on $\mstar^{\rm cen}$ for dwarfs, there is only about a $\rm 1-2\%$ chance for a central to have such a massive satellite, which is again roughly consistent with the results shown in Figure \ref{fig:ngtN}.

\subsection{Can they be detected?}
\label{sec:detect}

\begin{figure}
 \centering
 \includegraphics[scale=0.25]{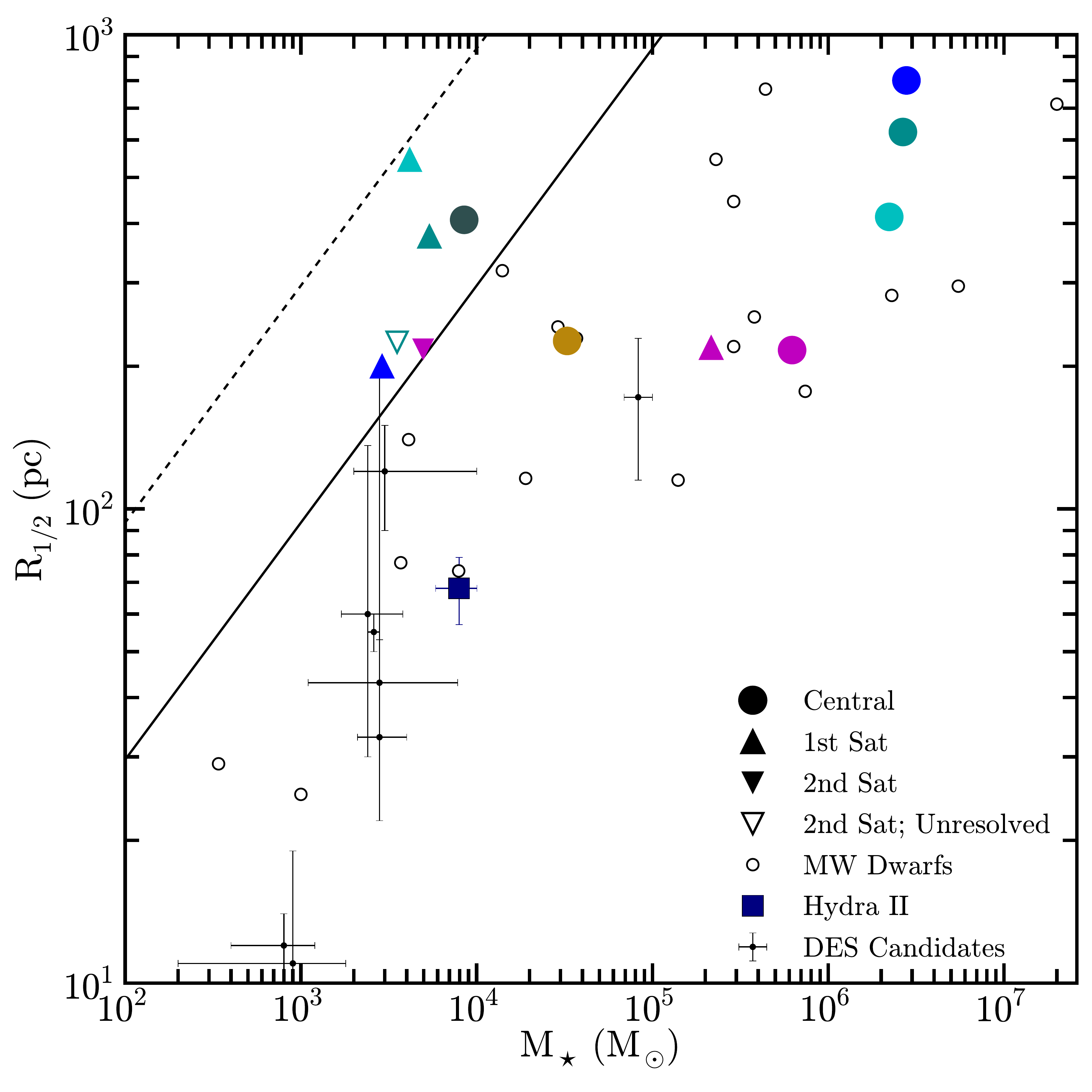}
 \caption{2D half-stellar-mass radii, $R_{1/2}$,  vs $\mstar$ for all resolved galaxies in our simulations (filled circles and triangles; colors are the same as in Figure \ref{fig:3panel}), the one unresolved satellite with $\mstar > 3000~\msun$ (open triangle), as well as for observed Milky Way dwarfs \citep[][open circles]{McConnachie:2012}, the newly discovered DES ultra-faint dwarf candidates \citep[][black points with error-bars]{DES2015}, and Hydra II \citep[][blue square]{Martin2015}. The solid line represents a surface brightness limit of $\rm 30~mag~arcsec^{-2}$ while the dashed line shows $\rm 32.5~mag~arcsec^{-2}$. Both lines and the Hydra II data point assume a stellar mass-to-light ratio of 1. Satellites as massive as ``dSph", which occur around $\sim 5\%$ of isolated dwarfs (see Figure \ref{fig:ngtN}), should be currently visible, but most of the satellites lie just out of reach, and will visible only with future surveys such as LSST or potentially even DES.
 \label{fig:rhalf_v_L}
}
\end{figure}

Our ultra-faint galaxies with $\mstar \simeq 3-30 \times 10^3~\msun$ all reside within low mass halos ($\mvir < 3 \times 10^9~\msun$).  They also have extremely diffuse stellar distributions, as expected for the ultra-faint, low dark-matter mass, ``stealth galaxies" discussed in B10. Figure \ref{fig:rhalf_v_L} shows the projected (2D) half stellar-mass radii, $R_{1/2}$, of our simulated galaxies vs their total stellar mass. The central galaxies are shown as colored circles and the satellites as triangles (the single unresolved satellite with $\mstar > 3 \times 10^3~\msun$ is shown as an open triangle). The open black circles show observed Milky Way dwarfs \citep{McConnachie:2012}. The black points with error bars are the half-mass radii and stellar masses of the eight recently reported ultra-faint satellite candidates from \citet{DES2015}, and the single blue square with error bars is Hydra II from \citet{Martin2015}. The surface brightness detection limit for SDSS is shown as a solid black line in the Figure.  It represents a constant peak central surface brightness for a Plummer profile
\begin{equation}
\Sigma_{\rm peak} = \frac{L}{\pi R_{1/2}^2}0.036~\lsun \rm pc^{-2}
\end{equation}
and corresponds to a surface brightness of $\rm \mu_V = 30~mag~arcsec^{-2}$ for solar absolute magnitude $\rm M_{\odot,V} = 4.83$ and assuming a stellar mass-to-light ratio of $\mstar/L = 1 (\msun/\lsun)$. 

Except for $\ufdone$, all of our simulated ``ultra-faint" dwarfs have surface brightnesses fainter than $\rm 30~mag~arcsec^{-2}$, and would qualify as ``stealth galaxies". \footnote{We have checked that the spatial radii of these galaxies are also well-resolved (not just their masses). Obviously, the plotted radii are much larger than our minimum force softening. A more demanding criterion is the  \citet{Power2003} radius which, for the dissipationless version of these runs, is $\gtrsim 100~\pc$ (see \citealt{Onorbe2015}). While this suggests that we are dynamically well resolved, the effect on the baryonic component within this radius is harder to determine. In lower resolution runs, the satellites' half-mass radii do vary by about a factor of two, but in both directions (some are larger, some smaller). Higher resolution simulations would be particularly useful for solidifying the expectations for the sizes of these galaxies.} The dashed line in Figure \ref{fig:rhalf_v_L} shows a surface brightness of $\rm 32.5~mag~arcsec^{-2}$, a limit that will likely be achieved by upcoming surveys such as the Large Synoptic Survey Telescope (LSST). Once the full co-added LSST data is collected, satellites this faint should be able to detected out to $\rm \sim 1~\mpc$ over half the sky \citep{Tollerud2008}.

In the more immediate future, the Dark Energy Survey (DES) is already online. DES will only be able to detect ultra-faint objects of this kind out to about $400~\kpc$, and is only slated to cover $\rm \sim 12\%$ of the sky over its first 5 years.  However, DES contains within its footprint the dwarf galaxy Phoenix. Phoenix has a stellar mass of $\mstar = 7.7 \times 10^5~\msun$ and is about $415~\kpc$ away. It is falling into the Milky Way with a velocity of $\rm \sim -100~\kms$ \citep{McConnachie:2012} and is likely on first approach, meaning that it is a good analog to the isolated systems simulated in this work. This suggests that Phoenix may be an excellent dwarf galaxy candidate to host an ultra-faint satellite detectable in the immediate future. It also has an advantage over more distant dwarfs like Cetus or Aquarius in that it is close enough that low luminosity satellites such as the ``ultra-faints" predicted in this work have a greater chance of being visible.

Although, according to Figure \ref{fig:ngtN}, there is only a $\rm \sim 35\%$ chance that Phoenix will host a subhalo massive enough to form a $\sim 3000 \msun$ satellite, once the volume that lies between us and Phoenix has been factored in, we find that the probability for finding an ultrafaint in its field increases to anywhere between $\rm 50\%-65\%$ according to the ELVIS simulations and depending on the specifics of abundance matching for $\mstar \sim 3000 \msun$ satellites.  In general, for future telescopes that have a limited number of pointings available, targeting the $\rm \sim 50~\kpc$ region around isolated dwarf galaxies should prove to be a much more efficient strategy than pointing into blank sky, as it should increase the chances of observing an ultra-faint satellite by $\sim 35\%$.

\section{Comparison to Previous work}
\label{sec:compare}

\citet{Sawala2014} run a series of 12 zoom-in simulations of Local Group-like environments at three levels of resolution, both with baryons and with dark matter only. They run each simulation with and without a cosmic UV background that turns on sharply at $z = 11.5$, and argue that the onset of the ionizing background radiation sets the mass scale at which all halos become dark. Their simulations shut down star formation in over $80\%$ of halos with present-day virial masses less than $10^9~\msun$. Their Figure 2 does seem to show convergence in their highest resolution simulation run with reionization, but the gas particle mass in their highest resolution run ($\sim 10^4~\msun$) is an order of magnitude higher than the total stellar mass formed in most of the the satellites that form in our simulations, indicating that they would fail to detect these galaxies. Furthermore, their reionization turns on at a higher redshift than does ours ($z = 10.65$), which may lead to a higher halo mass cutoff for star formation.

\citet{Shen2014} run a zoom-in simulation of a small group of seven dwarf galaxies with a range of virial masses ($4 \times 10^8~\msun$ $ \leq \mvir \leq 4 \times 10^{10}~\msun$). They find that galaxies only form in the halos with present day $\mvir > 10^9~\msun$, and of the halos that form galaxies, the two with $\mvir < 10^{10}~\msun$ only form stars long after the end of reionization. Although they do not model $\rm H_2$ cooling nor self-shielding in their simulations, they do approximate the effects of $\rm H_2$ cooling and run a parallel simulation without a cosmic UV background. In their reionization-free simulation, all dwarfs form stars, but they show that the gas in the three low mass halos never reaches the column density required for self shielding. Their baryonic particle mass is $\sim 10^3~\msun$, the same particle mass we achieve in the low resolution runs of $\dtwo$ (see \citealt{Onorbe2015} for details). In our low resolution runs, each of the most massive subhalos forms an object with stellar mass similar to its high resolution counterpart, but at this resolution each satellite object consists of only $3-5$ star particles. While it is difficult to say conclusively that resolution is the main driver of their apparent halo mass limit for star formation, in general it has been the case that every time resolution is increased in simulations, the minimum virial mass for a halo that can form stars has approached lower values \citep{Hoeft2006, Onorbe2015}. In addition to enabling the formation of galaxies with $\mstar$ lower than the low resolution baryonic particle mass, higher resolution simulations allow for a more accurate description of shielding, which will also affect the minimum halo mass that can form stars. Furthermore, resolving the formation of dense substructures that collapse under self-gravity and become self-shielding requires resolving the Jeans/Toomre mass of the galaxies, which can be as low as $\sim 1000~\msun$ (and corresponds to a required force/gravitational softening of at least $< 10-100~\pc$) in baryon-poor dwarfs.

Our simulations do not escape the challenges of imperfect resolution, but because our particle mass is an order of magnitude lower than most simulations that attempt to form galaxies in low mass halos, we can push down the predicted low mass limit for the halos that can form ultra-faint satellites. Rather than attempt to make a prediction for the low\textit{est} mass halo that can form a galaxy, we predict that ultra-faint dwarf galaxies \textit{can} form in dark matter subhalos as low mass as $\mpeak \sim 5 \times 10^8~\msun$. While we cannot state with certainty that these galaxies are fully converged in our high resolution runs, the stellar masses of the satellites do show a better convergence between low and high resolution runs than do the centrals. (see \citealt{Onorbe2015} for a discussion of the convergence of the centrals). Additionally, at these low masses, we still see sensitivity to spatial resolution and feedback (e.g. the second satellite of $\dmid$). More and higher resolution runs of dwarfs with $\mvir \sim 10^8 - 10^{10}~\msun$ that vary feedback prescriptions, reionization onset and spectrum, and particle mass are needed to fully probe the low mass end of the simulated stellar mass function and the $\mstar - \mhalo$ relation.

\section{Discussion and Conclusions}
\label{sec:dis}

In the \lcdm~ paradigm, all dark matter halos, from those around giant galaxy clusters to those hosting ultra-faint galaxies, should be filled with subhalos. We have used ultra-high resolution ($m^{\rm gas}_{p} \approx 255~\msun$) simulations with the PSPH version of \texttt{Gizmo} \citep{Hopkins2014b} and the Feedback in Realistic Environments (FIRE) prescriptions \citep{Hopkins2014a} to predict that subhalos of isolated dwarf halos should form galaxies. 

Most of the subhalos around dwarf galaxies are expected to be of low-mass ($\mpeak \lesssim 10^9~\msun$) and we predict that some should host ultra-faint galaxies with $\mstar \lesssim 10^4~\msun$.  If these tiny satellites are observed, it would provide evidence that dark matter substructure persists to very small scales, as predicted in the standard paradigm.  

Using the dark-matter-only simulations of the ELVIS suite, we show that each isolated field dwarf ($\mstar \sim 10^{6}~\msun$) galaxy in the Local Group has about a $\rm 35\%$ chance of hosting at least one satellite with $\mstar > 3000 \msun$. The extended $\rm \sim 50~\kpc$ regions around known field dwarfs in the Local Group should prove to be fruitful search areas for ultra-faint satellites, as each pointing towards a dwarf also contains all of the volume of the Milky Way dark matter halo along the line of sight to that region. The Phoenix dwarf galaxy in particular is an excellent target due to its proximity to the Milky Way and the high probability that it is on first infall. 

Although we consider only isolated dwarfs in this work, it is worth noting that some satellites of the Milky Way may also have their own satellites. \cite{Deason2015} show, using the ELVIS simulations, that approximately $7\%$ of Milky Way satellites (but as many as $25\%$ depending on the infall time and mass of the group), fell in as a part of LMC-size groups, and that several of the recent DES satellites are likely satellites of the LMC. A simple calculation for a rigid satellite assuming the common halo mass scale for Milky Way dwarf spheroidals ($\mhalo = 3 \times 10^{9}~\msun$) orbiting a Milky Way-size host ($\mvir = 10^{12}~\msun$) shows that Fornax could potentially hold on to ultra-faint satellites residing within its inner $17~\kpc$. Leo I and Leo II could each host ultra-faints out to $\sim 30~\kpc$, and Sculptor out to $10~\kpc$, assuming each satellite is on first infall. This suggests that these galaxies might also be suitable candidates to host ``satellites of satellites".

We do not see a sharp cut-off or break in the $\mstar - \mhalo$ relation, at least for $\mpeak > 5 \times 10^8~\msun$. We do, however, see a sharp cut-off in halos that host galaxies with uniformly ancient stellar populations. The ultra-faint dwarfs in our simulations form most of their stars in the first billion years after the Big Bang and are subsequently deprived of the cold gas required for star formation due to the ionizing background radiation, and not by infall into a more massive dark matter halo. We predict that below a critical mass threshold ($\mhalo \sim 5 \times 10^9~\msun$, $\mstar \sim 3 \times 10^4 \msun$) all galaxies are ubiquitously ancient and, unlike more massive galaxies ($10^7 < \mstar / \msun < 10^9$) that are nearly uniformly star forming in the field \citep{Geha:2012kx}, both central and satellite ultra-faint galaxies should all be quenched. 

For satellites with $\mstar > 10^5~\msun$, it is likely that infall time is one of the primary factors in quenching their star formation. Massive satellites ($10^{8.5} < \mstar / \msun< 10^{9.5}$) have been shown to have extremely long quenching timescales, probably due to the cutting off of the fresh gas supply after infall \citep{deLucia2012, Wetzel2013, Wheeler2014}. Slightly less massive galaxies ($10^6 < \mstar / \msun < 10^8$) are likely quenched over much shorter timescales by the tidal or ram-pressure forces they experience upon falling in as satellites \citep{SlaterBell2014, Fillingham2015}. However, it is likely that the quenching of ultra-faints is completely independent of their infall time, and set by the timing of the onset of and the mass they had at reionization.

The precise stellar masses of the ``ultra-faint" dwarfs in our simulations are likely also sensitive to the reionization redshift (with an earlier onset of reionization likely to quench star formation in higher mass halos) and may also be sensitive to star formation physics at very low metallicities. We plan to explore these dependencies in future work, but here we point out that it is possible to form ultra-faint satellites of regular dwarf galaxies and highlight that they are quenched by reionization rather than infall.

The recent discovery of up to nine Milky Way dwarf satellite galaxy candidates in the Southern Sky by DES \citep{DES2015, Koposov2015} has important consequences for our understanding of star formation in low mass galaxies. Of particular interest is DES J0344.3-4331 (Eridanus II), because of its large distance from the Milky Way ($> 330~\kpc$) \citep{DES2015}. At this distance, whether the satellite is on first infall or even headed out after a pericentric passage, it was likely accreted within the last $\sim 2~\gyr$. Given that the initial age estimate for Eridanus II shows that it likely has an ancient stellar population \citep[$\sim 10~\gyr$;][]{DES2015} \footnote{\citet{Koposov2015} suggest that Eridanus II could possibly have a population of young stars in addition to the old stellar population.}, this would mean that it was quenched long before infall. If confirmed in follow-up observations, Eridanus II would be the first known ultra-faint galaxy shown to be quenched in the field, supporting our findings that, below a critical mass scale $\sim 5 \times 10^9 \msun$, all galaxies host ancient stellar populations quenched by reionization-related feedback, and not by environmental processes. This would be a clear example of a reionization ``fossil", as first discussed by \citet{Ricotti2005}.

\section*{Acknowledgments} 
We thank M. C. Cooper for helpful discussions. This work used computational resources granted by NASA Advanced Supercomputing (NAS) Division, NASA Center for Climate Simulation, Teragrid and by the Extreme Science and Engineering Discovery Environment (XSEDE), which is supported by National Science Foundation grant number OCI-1053575 and ACI-1053575, the latter through allocation AST140080 (PI: Boylan-Kolchin). Support for this work was also provided by NASA through \textit{Hubble Space Telescope} theory grants (programs AR-12836 and AR-13888) from the Space Telescope Science Institute (STScI), which is operated by the Association of Universities for Research in Astronomy (AURA), Inc., under NASA contract NAS5-26555. CW and OE were supported by \textit{Hubble Space Telescope} grants HST-AR-13921.002-A and HST-AR-13888.003-A, and SGK by the National Science Foundation Grant AST-1009999. Support for PFH was provided by the Gordon and Betty Moore Foundation through Grant \#776 to the Caltech Moore Center for Theoretical Cosmology and Physics, an Alfred P. Sloan Research Fellowship, NASA ATP Grant NNX14AH35G, and NSF Collaborative Research Grant \#1411920. DK received support from National Science Foundation grant number AST-1412153, funds from the University of California, San Diego and XSEDE allocation TG-AST-120025.


\label{lastpage}
\end{document}